\documentclass[twocolumn,preprintnumbers,amsmath,amssymb,pre]{revtex4}

\usepackage{graphicx}
\usepackage{dcolumn}
\usepackage{bm}
\usepackage{amssymb}
\usepackage{wasysym}

\begin{document}

\title{Assortative mixing in Protein Contact Networks and protein folding kinetics}
\author{Ganesh Bagler} 
\email{ganesh.bagler@gmail.com}
\author{Somdatta Sinha\footnote{Corresponding Author}}
\email{sinha@ccmb.res.in}
\affiliation{Centre for Cellular and Molecular Biology, Uppal Road, Hyderabad, 500007 India.}
\date{\today}

\begin{abstract}
Starting from linear chains of amino acids, the spontaneous folding of
proteins into their elaborate three-dimensional structures is one of the remarkable
examples of biological self-organization. We investigated native state structures of
$30$ single-domain, two-state proteins, from complex networks perspective, to understand
the role of topological parameters in proteins' folding kinetics, at two length scales -
as ``Protein Contact Networks (PCNs)'' and their corresponding ``Long-range Interaction
Networks (LINs)'' constructed by ignoring the short-range interactions.

Our results show that, both PCNs and LINs exhibit the exceptional
topological property of ``assortative mixing'' that is absent in all
other biological and technological networks studied so far. We show
that the degree distribution of these contact networks is partly
responsible for the observed assortativity. The coefficient of
assortativity also shows a positive correlation with the rate of
protein folding at both short and long contact scale, whereas, the
clustering coefficients of only the LINs exhibit a negative
correlation. The results indicate that the general topological
parameters of these naturally-evolved protein networks can effectively
represent the structural and functional properties required for fast
information transfer among the residues facilitating
biochemical/kinetic functions, such as, allostery, stability, and the
rate of folding.

\textbf{Supplementary Information:} Supplementary data are available at \emph{Bioinformatics} online.\\
\end{abstract}


\maketitle


\section{Introduction}
Inside the cell, proteins are synthesized as linear chains of amino acids, which fold
into unique three-dimensional structures (`native states'). The wide range of
biochemical functions performed by the proteins are specified by their detailed
structures. Despite the large degrees of freedom, surprisingly, proteins fold into their
native states in a very short time, which is known as Levinthal's
Paradox~\cite{levinthal}. Although, given suitable conditions, some small proteins can
reach their native state in a single concerted step, many others fold in stages with
initial conformational events long before the final (`native') structure
appears~\cite{anfinsen_science}. Structural changes and chemical interactions occur
throughout the entire folding process, and strongly cooperative mechanisms are necessary
to bring the protein in its native conformation within a very short time
period~\cite{foldon_englander_PNAS2005}. The fast folding is a result of the catalytic
effect of the formation of clusters of residues in contact with each other, which have
high preferences for the early formation of secondary structures (helices, sheets, and
loops) in the presence of significant amounts of long-range tertiary structure
interactions~\cite{nolting_2000}.

The folding mechanism, kinetics, structure and function of proteins are intimately
related to each other. Misfolding of proteins into non-native structures can lead to
several disorders~\cite{taubes_Science1996}. Correlating sequence with structure, as
well as understanding of folding kinetics has been an area of intense activity for
experimentalists and theoreticians~\cite{fersht_book,branden_and_tooze_book}. Among the
different theoretical approaches used for studying protein structure, function, and
folding kinetics, the graph theoretical approach, based on perspectives from complex
networks, has been used recently to study protein
structures~\cite{cabios,protnet:PRE,protnet:JMB,protnet:Biophys,protnet:BaglerSinha,brinda_biophys2005,amitai_JMB2004,dokholyan_pnas2002,jung_lee_moon,rao_caflisch}.

It is known that folding mechanisms are largely determined by a protein's topology
rather than its inter-atomic interactions~\cite{baker99a}. With that understanding, we
build graph-theoretical models of protein structures to investigate various topological
properties at two different length scales, and study their possible role in the kinetics
of the protein folding. We use a coarse-grained complex network model of a protein
structure, viz. the Protein Contact Network (PCN), by ignoring the fine-grained
atomic-level details, and model the three dimensional structure as a system constituted
of amino acid units, put in place by noncovalent interactions. Long-range interactions
are known to play a distinct role in determining the tertiary structure of the
proteins~\cite{epand_1968}, as opposed to the short-range interactions, which could
largely contribute to the secondary structure formations. We consider the Long-range
Interaction Network (LIN) of each protein, which are subsets of the corresponding PCNs,
constructed by ignoring the short-range interactions. The idea behind studying LINs is
to understand the contribution of the long-range interactions to the topological
properties, and their correlation to a biophysically relevant property, viz. rate of
protein folding.

This study aims to address the question---Can general network parameters, derived from
native-state structures of proteins, uncover features about the relationship of the
structural properties to the folding kinetics of the proteins? To study this we choose
single domain, two-state folding proteins that belong to different structural
classes~\cite{SCOP} for which the kinetic parameter of rate of folding, ($k_F$) is
available. Our analysis of the coarse-grained network representations of protein
structures uncover the exceptional topological property of a high degree of assortative
mixing at both length scales (PCN and LIN) in these naturally-occurring, evolutionarily
selected, biological networks. Assortative mixing in LINs indicates that this feature in
PCNs is independent of short-range interactions. The coefficient of
assortativity~\cite{r:newman}, a measure of assortative mixing, are also found to be
considerably high for both PCNs and LINs. By constructing appropriate control networks,
we further demonstrate that the degree (connectivity) distribution of the PCNs alone can
partially account for the presence of assortativity in these networks.

To enumerate the contribution of these global parameters obtained from the
coarse-grained network model of protein structures to their biophysical properties, we
show that the coefficient of assortativity of PCNs and LINs tend to have positive
correlation with the experimentally determined rate of folding of these proteins. This
implies that assortative mixing, that tends to connect highly-connected residues to
other residues with many contacts, may assist in speeding up of the folding process. In
contrast, the average clustering coefficients of LINs show a good negative correlation
with the rate of folding, indicating that clustering of amino acids, that participate in
long-range interactions, into cliques, slows down the folding process. Interestingly,
the average clustering coefficients of PCNs show negligible correlation, thereby
implying that the short range interactions can reduce the negative effect on their
folding kinetics.

Three parameters--- CO (Contact Order)~\cite{co}, LRO (Long Range Order)~\cite{lro}, and
TCD (Total Contact Distance)~\cite{tcd} ---based on sequence distance per contact and/or
total number of contacts per residue of the proteins, have also been shown to have
negative correlation to their rate of folding~\cite{co,lro,tcd}. The accuracy of
prediction of the rate of folding, with parameters $LRO$ and $TCD$, remain unchanged if
short-range interactions are not included in the calculation. Here, along with
delineating the role of long-range interactions, we have attempted to show that general
network parameters, such as, clustering coefficient and assortativity, that are
widely-used in networks of diverse origins (technological, biological and social), can
not only give an insight into their structural properties, but can also be used as
indicators of specific biophysical processes, such as, of protein folding.

\section{Methods}
\subsection*{Construction of PCN, LIN, and their Random Controls}
The Protein Contact Network (\textbf{PCN}) was modeled from the native-state protein
structures as available in PDB~\cite{PDB}. The $C_{\alpha}$ atom of each amino acid was considered
a `node', and any two amino acids were said to be in spatial contact (`link') if there
existed a threshold distance ($R_c \le 8\AA$) between their $C_{\alpha}$ atoms.

The Long-range Interaction Network (\textbf{LIN}) of a PCN was obtained by considering,
other than the backbone links, only those `contacts' which occur between amino acids
that are `distant' (i.e. separated by $12$ or more amino acids) from each other along
the backbone. Thus formed, a LIN is a subset of its PCN with same number of nodes
($n_r$) but fewer number of links due to removal of the short-range contacts.

Two types of \textbf{random controls} were created for the PCNs of
the proteins. The polypeptide backbone connectivity was kept intact
in both the random controls, while randomizing the noncovalent
contacts. For every protein, $100$ instances of each type of random
control were generated from its PCN. Average of all the instances
were used as a representative of the parameters and properties, and
compared with that of the PCNs and their LINs.

\emph{Type I}: This random control network has the same number of residues ($n_r$) and
number of links/contacts ($n_c$) as those of the PCN, except that the contacts were
created randomly by avoiding duplicate and self contacts.

\emph{Type II}: Apart from maintaining the number of nodes ($n_r$) and contacts ($n_c$),
the connectivity distribution of PCNs was also conserved in this control network.
To ensure adequate randomization, the pattern of pair-connectivity was randomized $2000$ times.

The details of methods of construction with illustration is given in Supplementary Data.

\subsection*{Data}
Except for Fig.~\ref{fig:lc01}, all studies have been done on $30$
single-domain, two-state folding, globular proteins, whose
experimental rate of folding ($ln(k_F)$) are available. The data
include $5$ all-$\alpha$, $13$ all-$\beta$, and $12$ $\alpha\beta$
class of proteins. The natural logarithms of rate of folding
($ln(k_F)$) of these proteins vary between $-1.48$ and $9.8$ and
have a range for the time of folding ($1/k_F$) of the order of
$10^{5}$ seconds. Sizes ($n_r$) of these proteins range from $43$ to
$126$ amino acids. The structural data for these studies were
obtained from the Protein Data Bank~\cite{PDB}. The preliminary
network analysis (shown in Fig.~\ref{fig:lc01}) was done on $110$
proteins ($43<n_r<2359$) from the major structural classes, which
include the $30$ single domain proteins mentioned above.

\subsection*{Network parameters}
The following parameters were studied for the PCN, LIN, and their random controls.

\emph{Shortest Path Length and Characteristic Path Length}---Shortest path length
($L_{ij}$) between any pair of nodes $i$ \& $j$ is the number of links that must be
traversed between them by the shortest route. The average of all shortest path lengths,
known as `characteristic path length' ($L$), is an indicator of compactness of the
network, and is defined as~\cite{watts:nature},
$$ L = \frac{2 ~ \sum_{i=1}^{n_r-1}\sum_{j=i+1}^{n_r}L_{ij}}{n_r(n_r-1)},$$
where $n_r$ is the number of residues in the network.

\emph{Clustering Coefficient}---Clustering coefficient is the measure of
\emph{cliquishness} of the network. Clustering coefficient of a node $i$, $C_i$, is
defined~\cite{watts:nature} as the $C_i~=~2*n/k_i(k_i-1)$, where $n$ denotes the number
of contacts amongst the $k_i$ neighbors of node $i$. Average clustering coefficient of
the network ($C$) is the average of $C_i$s of all the nodes in the network and is
referred to as `clustering coefficient' unless specified otherwise.

\emph{Degree and Remaining Degree}---Degree~($k$) is defined as the total number of
neighbors a node is connected to. Degree is one of the measures of `centrality' of a
node in the network---the larger the degree more important it is. \emph{Remaining
degree} is one less than the total degree of a node~\cite{r:newman}. Other measures,
based on degree, are maximum degree, $k_{max}$, average degree, $\langle k \rangle$, and
the average degree of nearest neighbors, $\langle k_{nn}(k) \rangle$.

\emph{Assortative Mixing and Coefficient of Assortativity}---A network is said to show
assortative mixing, if the high-degree nodes in the network tend to be connected with
other high-degree nodes, and `disassortative' when the high-degree nodes tend to connect
to low-degree nodes. The Coefficient of Assortativity ($r$) measures the tendency of
degree correlation. It is the Pearson correlation coefficient of the degrees at either
end of a link and is defined~\cite{r:newman} as,
$$r=\frac{1}{\sigma_{q}^{2}} \sum_{jk}jk(e_{jk}-q_j q_k),$$
where $r$ is the coefficient of assortativity, $j$ and $k$ are the degrees of nodes,
$q_j$ and $q_k$ are the \emph{remaining degree} distributions, $e_{jk}$ is the joint
distribution of the remaining degrees of the two nodes at either end of a randomly
chosen link, and $\sigma_{q}$ is the variance of the distribution $q_k$.

\section{Results}
\subsection*{Clustering coefficients of PCNs and LINs}
\begin{figure}
\begin{center}
\includegraphics[scale=0.55]{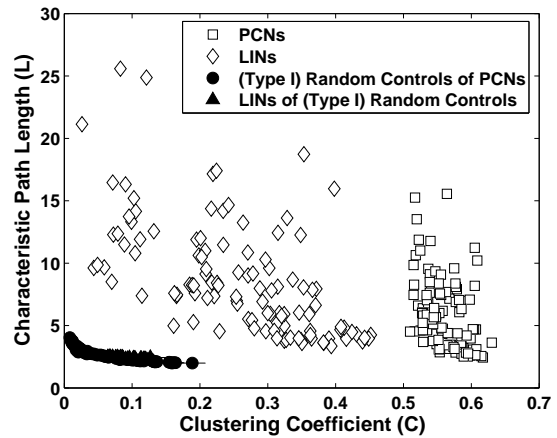}
\end{center}
\caption{\label{fig:lc01} L-C plot for 110 proteins from different
structural classes: PCNs ($\square$), LINs ($\Diamond$), Type I
Random Controls of PCNs ($\newmoon$) and LINs($\blacktriangle$).
Error-bars in the random controls data indicate standard deviations
in $L$ and $C$ for each protein computed over $100$ instances.}
\end{figure}

PCNs from a large set of proteins have earlier been
shown~\cite{protnet:PRE,protnet:JMB,protnet:Biophys,protnet:BaglerSinha}
to have high degree of clustering, which contributes to their
``small-world''~\cite{watts:nature} nature. To study if the PCNs
and their corresponding LINs of proteins have similar topological
properties such as, characteristic path length ($L$) and clustering
coefficient ($C$), we plotted the $L$ versus $C$ graph in
Fig.~\ref{fig:lc01} for 110 proteins from the four major structural
classes (i.e., $\alpha$, $\beta$, $\alpha+\beta$, and
$\alpha/\beta$). The figure also shows their corresponding Type I
random controls. The Type II random controls were found to be
indistinguishable from the Type I controls and not shown in
Fig.~\ref{fig:lc01}.

The results indicate two major differences between the topological
properties of the PCNs and their corresponding LINs. The PCNs of
these proteins have high clustering coefficients
($C_{PCN}=0.562\pm0.029$) compared to their random controls, whereas
the LINs show distribution in $C$ over a range
($C_{LIN}=0.259\pm0.109$), even though their random controls were
almost indistinguishable from those of PCNs. $L$ and $C$ of random
controls of PCNs were $2.621\pm0.411$ \& $0.0557\pm0.0476$ and that
of their LINs were $3.256\pm0.056$ \& $0.075\pm0.012$. The LINs also
have a little higher characteristic path lengths
($L_{LIN}=8.72\pm4.564$) than PCNs ($L_{PCN}=5.818\pm2.826$) owing
to their reduced number of contacts as compared to those in PCNs.
This indicates that the differences in $C_{LIN}$s may assign
specificity to the protein networks at this length scale, which is
otherwise lost with the short range contacts in PCNs, rendering the
generic property of high clustering and compactness. The role, if
any, the differential extent of clustering in the protein contact
networks at the two length scales may play  in their kinetics of
folding process is shown later.

\begin{figure*}
\begin{center}
\includegraphics[scale=0.90]{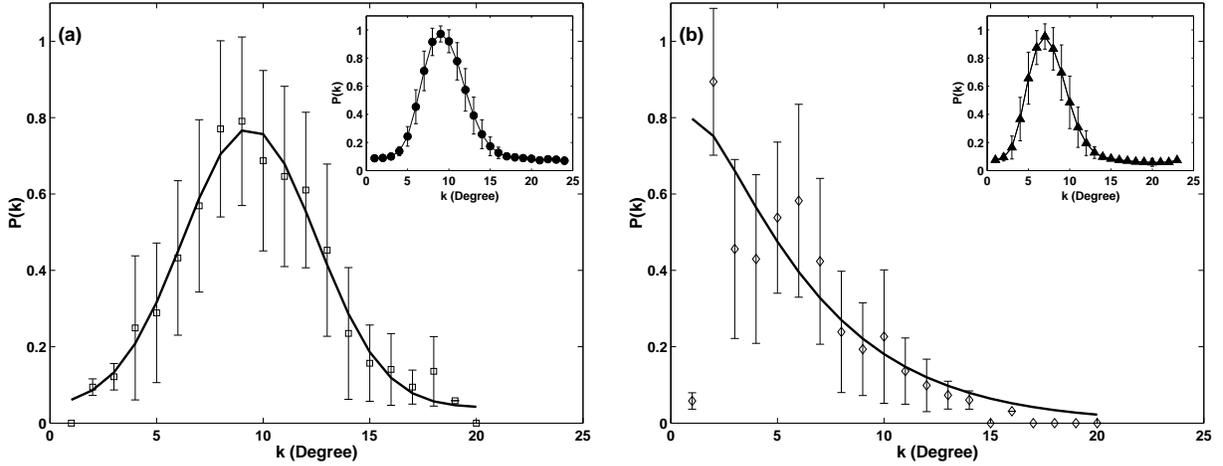}
\end{center}
\caption{\label{fig:degree_dist} Normalized degree
distributions~$P(k)$ of (a) PCNs and (b) LINs. Shown in the insets
are (a) Type I Random Controls of PCNs and (b) their LINs. Thick
lines are the best-fit curves for the means of the data. Error-bars
indicate standard deviation of the data for $P(k)$ of nodes with
degree $k$ across the $30$ proteins analyzed.}
\end{figure*}

\begin{figure*}
\begin{center}
\includegraphics[scale=0.90]{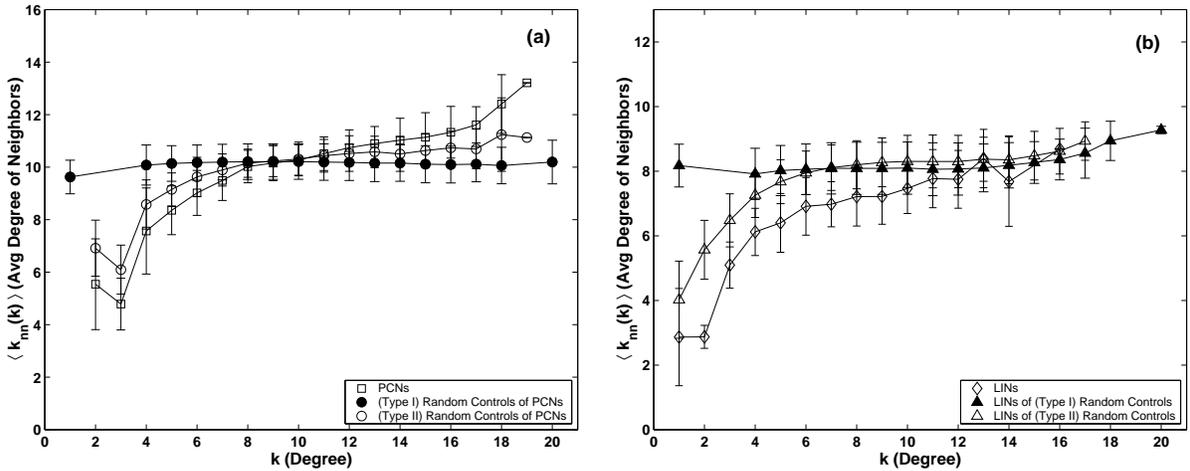}
\end{center}
\caption{\label{fig:degree_corr01} Degree correlation pattern for
(a) PCNs and (b) LINs. Assortative mixing of PCNs~($\square$) and
LINs~($\Diamond$) as compared to Type I Random Controls of
PCNs~($\newmoon$) and their LINs~($\blacktriangle$), and Type II
Random Controls of PCN~($\fullmoon$) and their
LINs~($\vartriangle$). Error-bars indicate standard deviation of the
data for $\langle k_{nn}(k) \rangle$ of nodes with degree $k$ across
the $30$ proteins and their controls.}
\end{figure*}

\subsection*{Degree distributions of PCNs and LINs}
The distribution of degrees in a network is an important feature, which reflects the
topology of the network, and is also a possible indicator of the processes by which the
network has evolved to attain the present topology. The networks in which the links
between any two nodes are assigned randomly have a Poisson degree
distribution~\cite{bollobas1981} with most of the nodes having similar degree.

Fig.~\ref{fig:degree_dist} shows the normalized degree distributions of PCNs and LINs of
the $30$ proteins studied. The frequencies of nodes were scaled with the largest degree
($k_{max}$) in the network (PCN or LIN) to obtain the $P(k)$ of a given protein, so that
proteins of different sizes can be compared. As seen in Fig.~\ref{fig:degree_dist}(a),
the PCNs have Gaussian degree distribution that best fits the equation
$$ y(x) = \frac{A}{w\sqrt{\pi/2}} \exp{\frac{-2(x-x_c)^2}{w^2}}$$ with $A=5.538$, $w=6.265$, and
$x_c=9.373$.

On the other hand, Fig.~\ref{fig:degree_dist}(b) shows that the degree distribution of
LINs is significantly different than those of PCNs. In LINs, most nodes were populated
in the low-degree region and very few of them have high degrees. The best-fit for the
LINs represent a single-scale exponential function~\cite{protnet:JMB}, $P(k) \sim
k^{-\gamma} \exp{(-k/k_c)},$ with $\gamma=0.24$ and $k_c=4.4$. The nodes of degree $1$
in LINs' degree distributions, are the N- and C-terminal amino acids that are at the
either end of the protein backbone. As expected~\cite{bollobas1981}, the Type I random
controls of the PCNs (Fig.~\ref{fig:degree_dist}(a), inset) have a Poisson degree
distribution. LINs of Type I random controls (Fig.~\ref{fig:degree_dist}(b), inset) too
have a Poisson degree distribution. The figure clearly shows that these properties are
the same for proteins irrespective of their functions and structural
classifications~\cite{protnet:JMB, protnet:BaglerSinha}.

\begin{figure*}
\begin{center}
\includegraphics[scale=0.70]{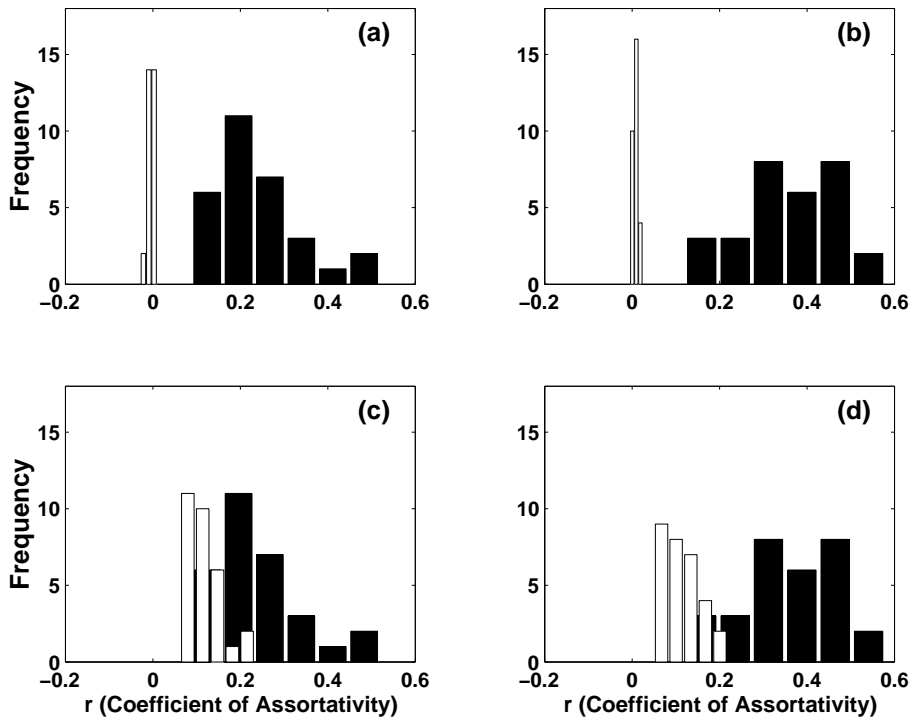}
\end{center}
\caption{\label{fig:degree_corr02} Histograms of `Coefficient of
Assortativity ($r$)' of PCNs, Type I and Type II Random Controls of
PCNs, and their LINs. (a) and (b) PCNs and LINs ($\blacksquare$) and
their (Type I) Random Controls ($\square$). (c) and (d) PCNs and
LINs  ($\blacksquare$) and their (Type II) Random Controls
($\square$).}
\end{figure*}

\subsection*{Assortative nature of PCNs and LINs}
The pattern of connectivity among the nodes of varying degrees can affect the
interaction dynamics in the network, and their degree correlation is used as a measure
to compute the strength and pattern of connectivity in a network. Average Degree of the
nearest neighbors, $k_{nn}(k)$, of nodes of degree $k$, is a parameter by which one can
measure and visualize the degree correlation pattern on a network. In the presence of
correlations, $k_{nn}(k)$ increases with increasing $k$ for an `assortative network',
and decreases with $k$ for a `disassortative network'~\cite{satorras:knn}.

Fig.~\ref{fig:degree_corr01} shows $\langle k_{nn}(k) \rangle$ versus $k$ plots for the
PCNs (a) and LINs (b) and the two types of random controls. The nature of the curves for
the PCNs ($\square$ in Fig.~\ref{fig:degree_corr01}(a)) and their LINs ($\Diamond$ in
Fig.~\ref{fig:degree_corr01}(b)) shows that both networks are characterized with
`assortative mixing', as the average degree of the neighboring nodes increased with $k$.
The curve shows a tendency to saturate at larger k - a feature that may be due to the
steric hindrance experienced by the connecting amino acids in the three-dimensional
structural organization of the protein. This steric hindrance restricts the position of
an amino acid in the three dimensional conformational space, and results in a maximum
values of degree ($k_{max}$) of a node. In comparison, the $\langle k_{nn}(k) \rangle$
remained almost constant for the Type I random control for both PCNs~($\newmoon$) and
LINs~($\blacktriangle$), indicating lack of correlations among the nodes' connectivity
in these controls.

The `coefficient of assortativity'~\cite{r:newman}, $r$, is a
global quantitative measure of degree correlations in a network, and
takes values as $-1 \leq r \leq 1$. $r$ is zero for no correlations
among nodes' connectivity, and takes positive or negative values for
assortative or disassortative mixing, respectively. The $r$ for both
PCNs and LINs of the $30$ proteins were found to be positive,
indicating that the networks are assortative.
Fig.~\ref{fig:degree_corr02} shows the histograms of $r$ of (a)
PCNs, (b) LINs, both in ($\blacksquare$), and their Type I random
controls ($\square$). The $r$ values of both PCNs as well as LINs of
all the proteins show significantly high positive values (range:
$0.09<r<0.52$ for PCNs, and $0.12<r<0.58$ for LINs) when compared to
other networks of diverse origins~\cite{r:newman}. Thus, the
networks modelling the native protein structures are clearly
characterized by high degree of assortative mixing at both short and
long contact scales. The Type I random controls in
Fig.~\ref{fig:degree_corr02}~(a \& b), for both PCNs and their LINs,
are distributed around zero, confirming the observation of lack of
degree correlations of the controls, made in
Fig.~\ref{fig:degree_corr01}.

These properties of positive $r$ and assortative degree correlations were also observed
(data not shown) for a large number of protein structures performing various cellular
functions and belonging to diverse structural categories (used
in~\cite{protnet:BaglerSinha}). This conclusively proves that the assortative mixing in
PCNs and LINs is a generic feature of protein structures. The role, if any, the assortative
nature of the protein contact networks at both length scales may play in their
kinetics of folding process is shown later.

\subsection*{Degree distribution partially accounts for assortativity}
To investigate whether the patterns of connectivity in the PCNs and LINs of the three
dimensional structures of the proteins contribute towards the observed assortativity, we
studied the assortative mixing and the `coefficient of assortativity' of Type II random
controls, in which the degree distribution of the PCNs were preserved  while randomizing
the pair-connectivities. Fig.~\ref{fig:degree_corr01}(c,d) show the degree correlation
plots of the Type II Random Controls of PCN~($\fullmoon$) and their
LINs~($\vartriangle$). It is clear that, unlike Type I random controls, the average
degree of the neighboring nodes increased with $k$ in Type II Random Controls, as seen
for the PCNs and LINs.

The histograms of the `coefficient of assortativity' ($r$) of Type II Random Controls
($\square$) are shown in Figs.~\ref{fig:degree_corr02} (c \& d). Here also, it can be
seen that the assortativity is partially recovered in the Type II random controls for
both PCNs and their LINs. Thus degree distribution partially explains the observed
assortative mixing. It implies that preserving the degree distribution of PCN, even
while randomizing the pair-connectivities, is important to partially restore the
assortative mixing in the random controls of PCNs as well as their LINs. The recovery of
assortative mixing in the LINs by Type II random controls of PCNs is even more
surprising, as the degree distribution of LINs (Fig.~\ref{fig:degree_dist}(b)) is very
different compared to the PCNs (Fig.~\ref{fig:degree_dist}(a)). This is especially
significant in the light of the observation~\cite{brunet_2004pre}
that one can rewire the links in a (scale-free) network to obtain assortativity or
disassortativity, to any degree, without any change in the degree distribution.

\subsection*{Correlation of protein network parameters to protein folding rates}

The general network parameters (e.g., $L$, $C$, and $r$) have been used to shed light on
the topology, growth and dynamics of widely different networks - physical, social and
biological. Here we show the relationship of these general topological parameters
(specifically, $C$ and $r$) obtained from our coarse-grained model of protein structures
(the PCNs and LINs), to a biophysical property underlying the
organization of the three-dimensional structure of the protein chains,
i.e., with the kinetics of protein folding. Below we have correlated
the available experimental data on the rate of folding of the
$30$ proteins with the two network parameters, $C$ and $r$ of the PCNs and their LINs.

\begin{figure}
\begin{center}
\includegraphics[scale=0.55]{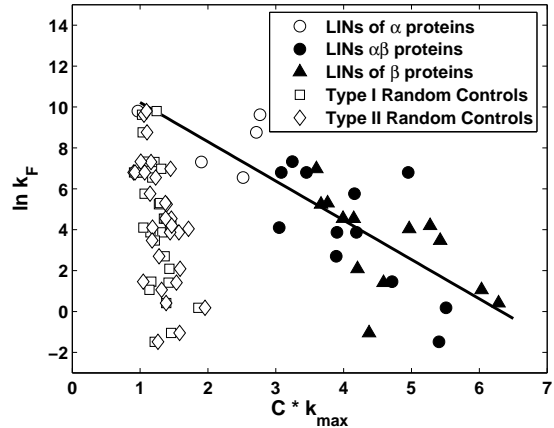}
\end{center}
\caption{\label{fig:lnkr_c01}
Rate of folding, $ln(k_F)$, has a
negative correlation, as indicated by the trendline, with Clustering Coefficient~($C$) LINs.}
\end{figure}

\subsubsection*{Average Clustering Coefficient and Rate of Folding}
Fig.~\ref{fig:lc01} shows that the PCNs and their LINs differ in
their clustering coefficients ($C$), with PCNs having similar but
high $C$, and their LINs having $C$ distributed over a range from
low to medium values. We did not find any significant relationship
between the clustering coefficient of the PCNs ($C_{PCN}$) and the
$ln(k_F)$ for all the $30$ proteins (correlation coefficient =
-$0.2437$; $p<0.2$). On the other hand, $ln(k_F)$ showed a high
negative correlation with the average clustering coefficient of the
corresponding LINs ($C_{LIN}$). Since the clustering coefficient
depends on the degree of the node, we plot, in
Figure~\ref{fig:lnkr_c01}, the $C_{LIN}*k_{max}$ with $ln(k_F)$ of
all the proteins. The plot shows significantly high negative
correlation (correlation coefficient = $-0.7712$; $p<0.0001$)
between the $C_{LIN}$s and the rate of folding for these
single-domain, two-state folding proteins. Fig~\ref{fig:lnkr_c01}
also shows that neither Type I nor Type II Random Controls show any
correlation with the rate of folding of the corresponding LINs.

$C_{LIN}$ enumerates number of loops of length three in the
Long-range Interaction Network. Thus $C_{LIN}$ essentially
correlates to the number of `distant' amino acids (nodes), those
separated by a minimum of $12$ or more other amino acids along the
backbone, brought in mutual `contact' with each other in the native
state structure of the protein. Understandably, more the number of
such long-range mutual contacts are required to be made in order to
achieve the native state, more is the time taken to fold, and hence
slower is the rate of folding. Interestingly, our result shows that
this feature is completely neutralised through the short-range
contacts in the PCNs. It may be mentioned that a comparable
correlation ($-0.7574$; $p<0.0001$) is observed between the Contact
Order ($CO$) of these 30 proteins with their $ln(k_F)$. It is
interesting to note that despite dissimilar quantities that $CO$ and
$C_{LIN}$ measure, the similar correlation coefficients essentially
indicate the important role of long-range contact formation in the
rate of folding.

\begin{figure}
\begin{center}
\includegraphics[scale=0.55]{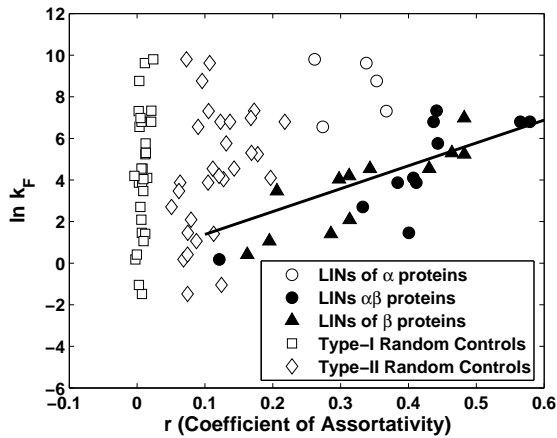}
\end{center}
\caption{\label{fig:lnkr_r01}
Positive correlation between the rate of
folding, $ln(k_F)$, and the Coefficient of Assortativity ($r$)
of LINs. The trendline is also shown.}
\end{figure}

\subsubsection*{Coefficient of Assortativity and Rate of Folding}
Unlike the clustering coefficients, the protein networks show high
coefficient of assortativity ($r$) at both length scales (i.e., for
the PCNs and their LINs). In Fig.~\ref{fig:lnkr_r01}, the rate of
folding of the proteins are plotted as a function of the coefficient
of assortativity of their LINs. There is an increasing trend of
$\ln(k_F)$ with increase in $r$. The five $\alpha$ proteins, all
having high rate of folding, do not follow the trend very well. The
correlation coeff.\ between the rate of folding ($ln(k_F)$) and $r$
of their LINs, excluding the five $\alpha$ proteins, is $0.6981$
($p<0.0005$). The same for the PCNs is calculated to be $0.5943$
($p<0.005$). The result implies that, along with showing assortative
mixing, the PCNs and their LINs both show significant positive
correlations with the rate of folding. Thus, the generic property of
assortative mixing in proteins tends to contribute positively
towards their kinetics of folding, and is fairly independent of the
short- and long-range of interactions. Here also the Type I Random
Controls, due to their coefficient of assortativity being clustered
around zero (Fig.~\ref{fig:degree_corr02}(b)), do not show any
correlation with the rate of folding. As is expected from
Fig.~\ref{fig:degree_corr01} and~\ref{fig:degree_corr02} , the
Type II random controls, on the other hand, are scattered owing to
the partial gain in assortativity, though they do not show any
definite trend with the rate of folding.

\section{Discussion}
In recent years, much interest is seen in the study of structure and dynamics of
networks, with application to systems of diverse origins such as, society, technology,
and biology etc~\cite{reka:thesis,dorogovtsev:book}. The aim of these studies has been to identify
the common organizational principles within these wide variety of systems, and identify
general network parameters that can correlate to the structure, function, and evolution
of each of the specific processes. Of these, biological networks are of special interest
as they are products of long evolutionary history. The protein contact network is
exclusive among other intra-cellular networks for their unique method of synthesis as a
linear chain  of amino acids, and then folding into a stable three-dimensional structure
through short- and long-range contacts among the residues. In this study,
our aim is to understand if the general network parameters can offer any clue to the
biophysical properties of the existing three dimensional structure of a protein, thereby
reflecting the commonalities in network organization in general.

Our coarse-grained complex network model of protein structures uncovers, for the first
time in a naturally evolved biological system, the interesting, and exceptional
topological feature of assortativity at both short and long length scale of contacts.
The assortative nature is found to be a generic feature of protein structures. We show
that the assortativity positively correlates to the folding mechanisms at both length
scale. This feature corroborates the known fact that the folding mechanisms are largely
independent of the finer details of the protein structure~\cite{baker99a}. Since
strongly cooperative mechanisms are necessary to bring the protein in its native
conformation within a very short time~\cite{foldon_englander_PNAS2005}, we have shown
that assortative mixing contributes positively towards speeding up the folding process
at different contact-length scales. The generality of assortative mixing in PCNs assume
greater importance in the light of the debate on whether protein folding kinetics is
under evolutionary
control~\cite{mirny_shakh_PNAS_1998,ScalleyKim_Baker_JMB_2004,Larson_etal_JMB_2002}.
Given the genetic basis and mode of formation of protein chains, the signature of
assortativity as an indicator to the rate of folding is clear.

We also delineate the difference in the property of clustering of
the nodes in the native structure at short and long length scales.
The PCNs have high degree of clustering, which contributes to their
`small-world' nature helping in efficient and effective dissipation
of energy needed in their
function~\cite{protnet:Biophys,protnet:BaglerSinha}. Our results
show that, in contrast, the corresponding LINs have significantly
lower and distributed clustering coefficients~(Fig.~\ref{fig:lc01}),
and they show a negative correlation with the rate of folding of the
proteins~(Fig.~\ref{fig:lnkr_c01}). This indicates that clustering
of amino acids that participate in the long-range interactions, into
`cliques' can slow down the folding process - possibly due to the
backbone connectivity and steric factors. However, the clustering
coefficient of PCNs \emph{do not} have any significant correlation
to the rate of folding, clearly indicating that the short-range
interactions may be playing a constructive and active role in the
determination of the rate of the folding process by reducing the
negative contribution of the LINs. Our results thus show that the
separation of the types of contacts in the PCNs and LINs clearly
delineate the length scale of contacts that play crucial role in
protein folding. It was recently shown that the $CO$ of the
Transition State Ensemble ($TSE$) is highly correlated to that of
their native state structure, and they both correlate equally well
with their rate of folding~\cite{paci2005}. This has been
attributed to the fact that the long-range contacts are mainly
located in the structural core that are formed early in the folding
process, and the formation of such contact networks leads to the
inverse correlation with the folding rates. Our results with general
parameters of the long-range interaction networks ($C_{LIN}$ and
$r_{LIN}$) corresponding to the native PCNs also reflect the crucial
role that long-range interactions play in their rate of folding.

After the synthesis in the cell, folding of the amino acid chain is important for
attaining the structure required to reach a functional state as soon as possible. This
happens through inter-residue non-covalent interactions at many length and time scales.
The folded structure have to confer stability, regions for binding of ligands of
specific shapes and sizes, transmit the information of binding/unbinding to other parts
of the protein, scaffold for retaining the functional regions along with the shape
suitable for the protein function. It is likely that many of these properties may
require opposing features to operate at different time and space scales. For example,
the `small-world' nature (high clustering) in the native protein structure is useful in
inter-residue signalling required for its function on binding and allostery. On the
other hand, the long-range interaction network have reduced clustering, which may
facilitate communication among distant residues in the native structure to some extent,
but such a feature can also increase the folding time as it requires distant residues in
the chain to come closer during the folding process. Thus, the evolved native structure
of the proteins show differential levels of clustering at two length scales. The
assortative mixing, on the other hand, helps in enhancing the folding process at both
length scales.

A large number of networks of diverse origin have been found~\cite{r:newman} to be of
disassortative nature, and questions regarding the origin of this property and whether
this is an universal property of complex networks, has been adjudged as ``one of the ten
leading questions for network research''~\cite{EPJB:round_table}. Our discovery of
assortativity in the amino acid networks in protein structures at short and long contact
scales questions the invoked generality of the property in natural networks. The
assortative nature of the social networks has been claimed to be originating from their
unusually high clustering coefficients and community structure~\cite{newman:society01}.
In proteins, LINs have high assortativity without necessarily having high clustering
coefficients. It would be interesting to study if the secondary structures provide any
role in shaping the ``community structure'' in these molecular networks that help in
conferring assortative mixing at both contact length
scales~\cite{newman:society01,community_vicsek_nature}.

Disassortative mixing observed in certain biological networks (metabolic signaling
pathways network, and gene regulatory network) is conjectured to be responsible for
decreasing the likelihood of crosstalk between different functional modules of the cell,
and increasing the overall robustness of a network by localizing effects of deleterious
perturbations~\cite{maslov:science2002}. In contrast to these two networks, PCNs are not
disassortative. For the PCN, one may put forward the possibility of the backbone chain
connectivity as a means of conferring greater robustness against perturbations.

From computational studies, it has been observed~\cite{r:newman,brunet_2004pre} that
assortative networks percolate easily, i.e., information gets easily transferred through
the network as compared to that in disassortative networks. Protein folding is a
cooperative phenomenon, and hence, communication amongst nodes is essential, so that
appropriate noncovalent interactions can take place to form the stable native state
structure~\cite{foldon_englander_PNAS2005}. Thus percolation of information is very much
essential and could lead to the observed cooperativity and fast folding of the proteins.
Hence assortative mixing observed in proteins could be an essential prerequisite for
facilitating folding of proteins.

\section*{Acknowledgements}
GB thanks CSIR, India for the Fellowship.
GB is grateful to Cosma Shalizi, Ch.\ Mohan Rao, and R. Sankaranarayanan for valuable suggestions.


\end{document}